\documentclass[twocolumn,superscriptaddress]{revtex4-1}

\usepackage{times}
\usepackage{graphicx}	% Include figure files

\usepackage{color}

\usepackage{soul} %needed for strikeout
\usepackage{amsmath}
\usepackage{hyperref}
\usepackage{physics}

\begin{document}

\preprint{accepted in Nature Photonics}

\title{Realisation of superabsorption by time reversal of superradiance}
\author{Daeho Yang}
\author{Seung-hoon Oh}
\author{Junseok Han}
\author{Gibeom Son}
\author{Jinuk Kim}
\affiliation{Department of Physics and Astronomy \& Institute of Applied Physics, Seoul National University, Seoul 08826, Korea}
\author{Junki Kim}
\affiliation{Department of Electrical and Computer Engineering$\cdot$Duke University, Durham, North Carolina 27708, USA}
\author{Moonjoo Lee}
\affiliation{Department of Electrical Engineering, Pohang University of Science and Technology(POSTECH), Pohang 37673, Korea}
\author{Kyungwon An}
\email{kwan@phya.snu.ac.kr}
\affiliation{Department of Physics and Astronomy \& Institute of Applied Physics, Seoul National University, Seoul 08826, Korea}

\date{\today}

\begin{abstract}
{\bf Emission and absorption of light lie at the heart of light-matter interaction\cite{MM_photodetection_review10}. 
Although emission and absorption rates are regarded as intrinsic properties of atoms and molecules, various ways to modify these rates have been sought in such applications as quantum information processing\cite{QI_quantum_interface_rmp10}, metrology\cite{SR_clock16} and light-energy harvesting\cite{LH_SR_prb08}. One of the promising approaches is to utilise collective behaviour of emitters as in superradiance\cite{SR_Dicke}. Although the superradiance has been observed in diverse systems\cite{SR_clock16, SR_HF, SR_rev82, SR_ion15, SR_superconducting14, ML_SR}, its conceptual counterpart in absorption has never been realised\cite{SA_theory14} to date.
Here, we demonstrate `superabsorption', the enhanced cooperative absorption, by implementing a time-reversal process of superradiance. The observed superabsorption rate is much higher than that of ordinary absorption while
the number of absorbed photons scales with the square of the number of atoms, exhibiting the cooperative nature of superabsorption. The present superabsorption, performing beyond the limitation of the conventional absorption, can facilitate weak-signal sensing\cite{MM_photodetection_review10}, light-energy harvesting\cite{SA_theory14} and light-matter quantum interfaces\cite{QI_quantum_interface_rmp10}.}
\end{abstract}
\maketitle

Dicke superradiance\cite{SR_Dicke} refers to the enhanced radiation of correlated emitters.
Unlike ordinary spontaneous emission of individual emitters, the correlation among emitters effectively makes them behave as a single giant dipole and thus the radiation intensity scales nonlinearly with the number of emitters.
In the conventional superradiance, the correlation slowly builds up via spontaneously emerging entanglement of close-spaced emitters through radiative decay initiated by vacuum fluctuations\cite{SR_HF,SR_rev82}.
The correlation buildup should be faster than its dephasing in order to achieve superradiance as in the latest experiments performed with Bose-Einstein condensates\cite{SR99_BEC}, trapped atoms\cite{SR15_cavity_modified}, and nitrogen-vacancy centres in diamond\cite{SR18_diamond_trillion}.

The recent technical advances in precision control of emitters make it possible to prepare a superradiant state by actively imposing quantum correlation without relying on the spontaneous emergence of correlation\cite{SR_clock16, SR_superconducting14, ML_SR}.
In experiments with superconducting circuits\cite{SR_superconducting14}, trapped ions \cite{SR_ion15} and atoms\cite{np16_trapped_atom_rempe} with high degree of control over individual qubits,
internal states of individual emitters were manipulated to observe subradiance as well as superradiance although the maximum number of emitters was limited to two.
In macroscopic ensembles of quantum dots\cite{SR16_sinlge_photon_QD} and cold atoms\cite{SR16_sinlge_photon_cold}, direct correlation imprinting was achieved by adopting a single-photon superradiance scheme \cite{SR_09SS_Scully} with a moderate state-preparation efficiency.
A superradiant state made of a large number of emitters with imposed correlation can generate the enhanced radiation at high efficiency without a correlation buildup delay. This feature is advantageous in superradiant light sources\cite{SR_clock16} as well as in robust quantum communication with efficient retrieval of stored photons\cite{SR_comm_exp}.

Pushing the envelope further, one can envisage the superabsorption, enhanced absorption by correlated emitters as the opposite of superradiance. 
Many studies suggest possibilities of increasing absorption efficiency by utilising collective behaviour of emitters in light-harvesting\cite{LH_SR_prb08, photosynthesis_nat17_review}, but most of them focus on enhancing excitation energy transfer with a single-photon superradiant state\cite{LH_SR_prb08} or through superradiance transition\cite{LH_jpcc12_superradiant_transition}, not directly seeking correlated absorption. 
Since absorption and emission of light are in the relation of time-reversal symmetry, 
the absorption rate of the so-called bright state responsible for superradiance also nonlinearly scales with the number of emitters. 
The bright state can be accessed in the superradiance ringing process\cite{Rabi_prl83,ringingSR_prx17}, but the atoms and the emitted field are highly entangled without a preferred phase and thus superabsorption of an externally introduced field does not occur.
The recent theoretical studies suggest to suppress the vacuum density of states for the downward transition of superradiant states and thereby allow only the upward transition for superabsorption\cite{SA_theory14}, but its experimental realisation is yet to be achieved.

\begin{figure*}
\centering\includegraphics[width=0.75\textwidth]{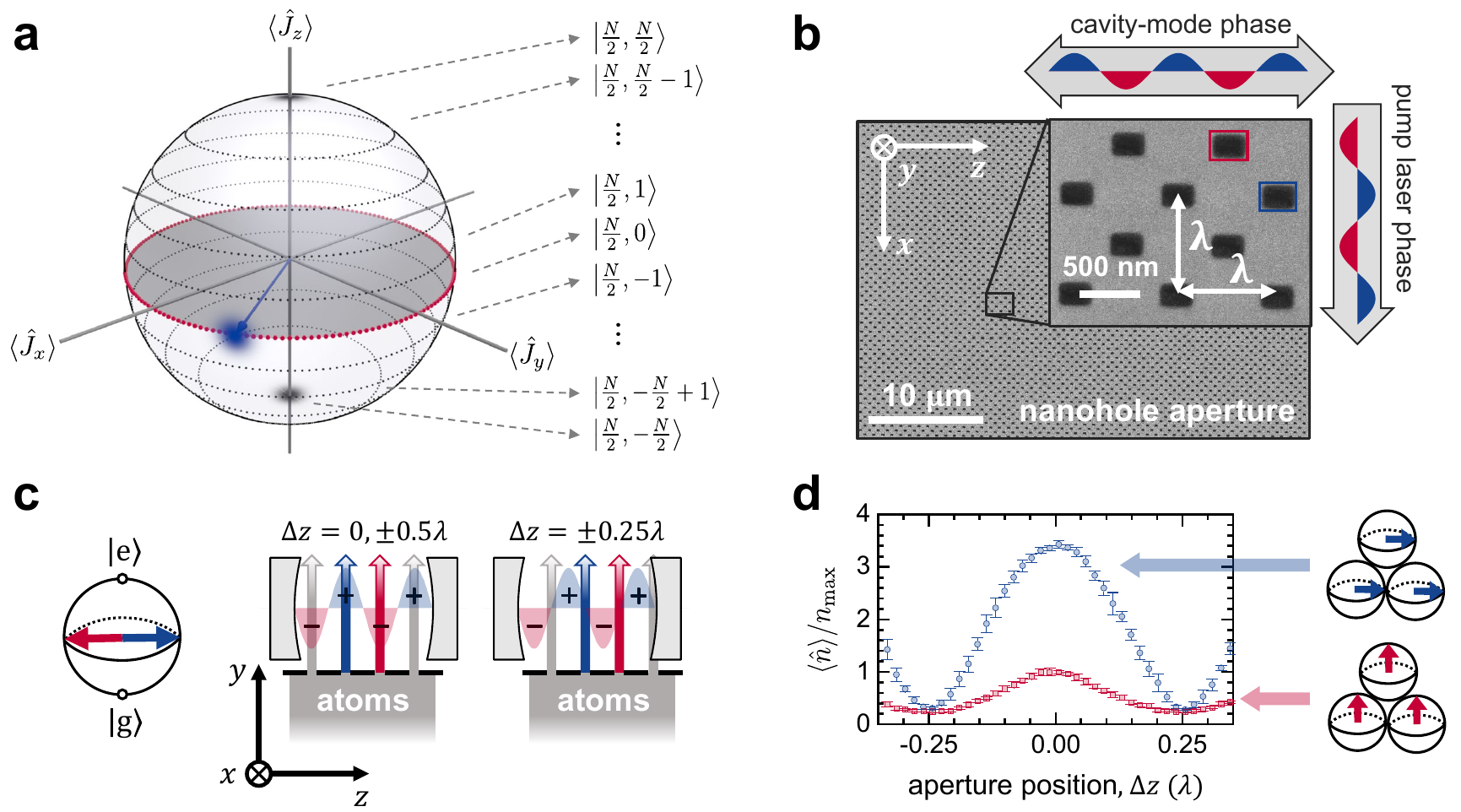}
\caption{
		\textbf{Experimental scheme and state preparation with a nanohole array. }
		{\bf a}, Dicke states $\ket{J,M}$ are represented as latitudinal rings (corresponding to the maxima of their Q-functions mapped on the Bloch sphere) in the $N$-atom Bloch sphere. Our superradiant state is localised in the equator like a spin coherent state. 
		{\bf b}, Images of the nanohole-array aperture captured with a scanning electron microscope. Two-level atoms traveling in the $y$ direction are excited by a pump laser propagating in the $x$ direction to a superposition state after filtered by the nanohole array aperture. The state-prepared atoms interact with the cavity mode, whose axis lies in the $z$ direction. Period of the nanohole array matches the transition wavelength of the two-level atom($\lambda$=791 nm of barium-138). The total array size is $50\lambda\times50\lambda$. The single(double)-headed arrow indicates the traveling(standing) wave nature of the pump(cavity) field with the sinusoidal oscillation depicting the phase variation. 
		{\bf c}, Bloch vector of the atoms passing through the red(blue)-marked hole is associated with a pump-laser phase of $0$($\pi$). Since the atom-cavity coupling of the atoms going through the red-marked hole is opposite in sign to that of the blue-marked hole, these two group of atoms experience the same atom-cavity relative phase. 
		{\bf d}, Experimentally observed radiation intensity as a function of $\Delta z$ is shown in a normalised average photon number $\expval{\hat{n}}/n_{\rm max}$ with $n_{\rm max}$ the maximum number of photons generated by fully excited atoms. The atoms in the same superposition state radiate much more than those in the fully excited state due to the correlation among the atoms. The mean number of atoms in the cavity is $N=2.7$ for both cases. Error bars indicate standard deviations from repeated measurements  (see Supplementary Note 10 for details).  
%		\comment{[280words]}
		}
\label{fig:scheme}
\end{figure*}

Here, we experimentally demonstrate the superabsorption by implementing a time reversal process of superradiance with phase control of a superradiant state in a cavity.
We reverse the whole system in time, not only the atomic state but also the photonic state.
In our scheme, in the absence of an input field, the superradiant state of correlated atoms would generate a coherent-state superradiant field\cite{ML_SR} with its phase determined by that of the atomic superradiant state. 
By controlling the phase of the prospective superradiant field opposite to that of an input field into the cavity, we can make the superradiant state undergo upward transition in the Dicke ladder, resulting in a time reversal of the superradiance, {\em i.e.}, the realisation of superabsorption. We observed greatly enhanced absorption with the number of absorbed photons scaling with the square of the number of atoms while its rate much greater than that of the ordinary ground-state absorption.

The superradiant state we consider is an $N$-atom symmetric state analogous to a spin coherent state, localised in the equator in the $N$-atom Bloch sphere (see Fig.~\ref{fig:scheme}{\bf a}), which is defined in terms of the collective spin operator $\hat{J_\mu}=\frac{1}{2}\sum^{N}_{i=1}\hat{\bf \sigma}_i^{\mu}$ with $\hat{\sigma}_i^\mu$ the Pauli matrices for the $i$th atom and $\mu=x, y, z$. 
In the absence of an input field, the emission process of the superradiant state $\ket{\Psi}_{\rm a}$ in a cavity can be approximately described by
\begin{equation}\label{eq:SR}
	\hat{U}(t)\ket{\Psi}_{\rm a}\ket{0}_{\rm f} = \ket{\Psi'}_{\rm a}\ket{\alpha}_{\rm f},
\end{equation}
where $\hat{U}(t)$\;$\equiv$\;$e^{-i\hat{H}t/\hbar}$ denotes the time-evolution operator of the Tavis-Cummings Hamiltonian $\hat{H}$\;$=$\;$\hbar g \sum_{i=1}^{N} \left( \hat{a}^\dagger \hat{\sigma_i} + \hat{a} \hat{\sigma_i} ^\dagger \right)$ with $g$ the atom-cavity coupling constant, $\hat{a}(\hat{a}^\dagger)$ the annihilation(creation) operator for the cavity field and $\hat{\sigma_i}(\hat{\sigma_i}^\dagger)$ the lowering(raising) operator for the $i$th atom, $\ket{0}_{\rm f}$ denotes a photonic vacuum state, $\ket{\alpha}_{\rm f}$ represents a photonic coherent state with an amplitude $\alpha$ and $\ket{\Psi'}_{\rm a}$ is the resulting atomic state by the time evolution. 
Equation (\ref{eq:SR}) describes a superradiance process\cite{ML_SR}, where the average photon number of the coherent state $\ket{\alpha}_{\rm f}$ is proportional to $N^2$. 
 The approximation used in Eq.~(\ref{eq:SR}) is valid under the condition $g\tau\ll 1$, where $\tau$ is the atom-field interaction time for the atoms traversing the cavity. 
The final state $\ket{\Psi'}_{\rm a}$ is a state rotated downward by an angle in the order of $(g\tau)^2$ from the initial state $\ket{\Psi}_{\rm a}$ in the $N$-atom Bloch sphere (see Supplementary Notes 6 \& 7 for derivations).
By introducing a field-phase-flipping operator $\hat{R}_{\pi}$ corresponding to $\pi$-rotation in the field phase space and by utilising the relation $\hat{R}_{\pi}\hat{U}(t)\hat{R}_{\pi}^\dagger=\hat{U}(-t)$, one can then show (see Methods for details)
\begin{equation}\label{eq:SA}
	\hat{U}(t)\ket{\Psi'}_{\rm a}\ket{-\alpha}_{\rm f}\simeq\ket{\Psi}_{\rm a}\ket{0}_{\rm f}.
\end{equation}
What it means is as follows. By preparing the cavity with an initial state $\ket{-\alpha}_{\rm f}$, we can achieve the time reversal of the superradiance process of Eq.~(\ref{eq:SR}). The atomic state returns to the initial state and the cavity field is completely absorbed to be the vacuum state with the absorbed number of photons proportional to $N^2$. 
Since $\ket{\Psi'}_{\rm a}$ is very close to $\ket{\Psi}_{\rm a}$ as discussed above, we can employ the superradiant state $\ket{\Psi}_{\rm a}$ instead of $\ket{\Psi'}_{\rm a}$ in the experiment to be described below.
Incidentally, time reversal idea was employed in coherent perfect (ordinary) absorption as time reversal of lasing\cite{CPA_Theory_PRL10}. 
Time-reversed photon pulse shaping was also used in efficient absorption of optical as well as microwave photons\cite{Ab_microwave14}.

In our experiment, a beam of two-level atoms (barium-138 with $^1$S$_0\leftrightarrow^3$P$_1$ transition at $\lambda$=791nm) traveling in the $y$ direction in Fig.~\ref{fig:scheme}{\bf b} goes through a nanohole-array aperture\cite{ML_vac} in a checkerboard pattern with a period of $\lambda$. Just behind the nanohole array, the atoms are excited by a pump laser propagating in the $x$ direction with a pulse area of $\Theta$ and then enter a high-finesse Fabry-P\'{e}rot cavity. By the nanohole array, the atomic vertical ($x$) position is localised at $\pi$-different equiphase planes of the pump laser and its horizontal ($z$) position localised at one of the cavity anti-nodes (Fig.~\ref{fig:scheme}{\bf c}). With both the atom-cavity coupling and the pump laser phase alternating their signs with a period of $\lambda/2$, every atom is then excited to a superposition state of the ground and excited states with the same atom-cavity relative phase\cite{ML_SR}. Under this condition, the atomic state can be written as $\ket{\Psi}_{a}=\prod_{k=1}^{N}\left[\cos(\Theta/2)\ket{\rm g}_k+\exp(-i\phi_0)\sin(\Theta/2)\ket{\rm e}_k\right]$, which for $\Theta=\pi/2$ corresponds to a superradiant state localised in the equator of the $N$-atom Bloch sphere. Here g(e) stands for the ground(excited) state of the atom and $\phi_0$ is a common atom-cavity relative phase. Our experiment belongs to the strong coupling regime for single atoms with $(\gamma_{\rm a}, \gamma_{\rm c}, \bar{g})$ = $2\pi \times (25, 131, 256)$kHz, where $\gamma_{\rm a}$($\gamma_{\rm c}$) is the atomic{(cavity-field) half linewidth} and $\bar{g}$ is the spatial-averaged (over the nanohole opening) atom-cavity coupling constant. Due to the strong coupling and a short atom-cavity interaction time $\tau\approx$100ns, the spontaneous emission into the free space and thus the decoherence of the atomic state can be neglected during the interaction time $\tau$.

Contrary to randomly distributed atomic dipoles, every dipole moment ${\bf p}$ of our atomic superposition state is the same with an identical phase and thus the total dipole moment in the cavity behaves like a macro dipole, $N{\bf p}$. Although each dipole resides in the cavity only for the short interaction time $\tau$, since the dipole moment of a newly entering atom is perfectly in phase with the preceding ones, coherence is maintained among the dipoles with different entering times. The sustained coherence among entering and exiting dipoles allows us to simplify our system as a macro dipole $N{\bf p}$ stationary in the cavity,
so the description by Eqs.~(\ref{eq:SR}) and (\ref{eq:SA}) can be applied to our experiment. As a supporting data, enhanced radiation or superradiance by  such $N$ atoms in a superradiant state is observed in the steady state (Fig.~\ref{fig:scheme}{\bf d}) without any input field. Due to the correlation among the dipoles, the superposition state($\Theta=\pi/2$) emits more photons than the fully excited state($\Theta=\pi$) when the holes of the aperture are centred at the cavity anti-nodes($\Delta z=0$). 

\begin{figure}
\includegraphics[width=0.5\textwidth]{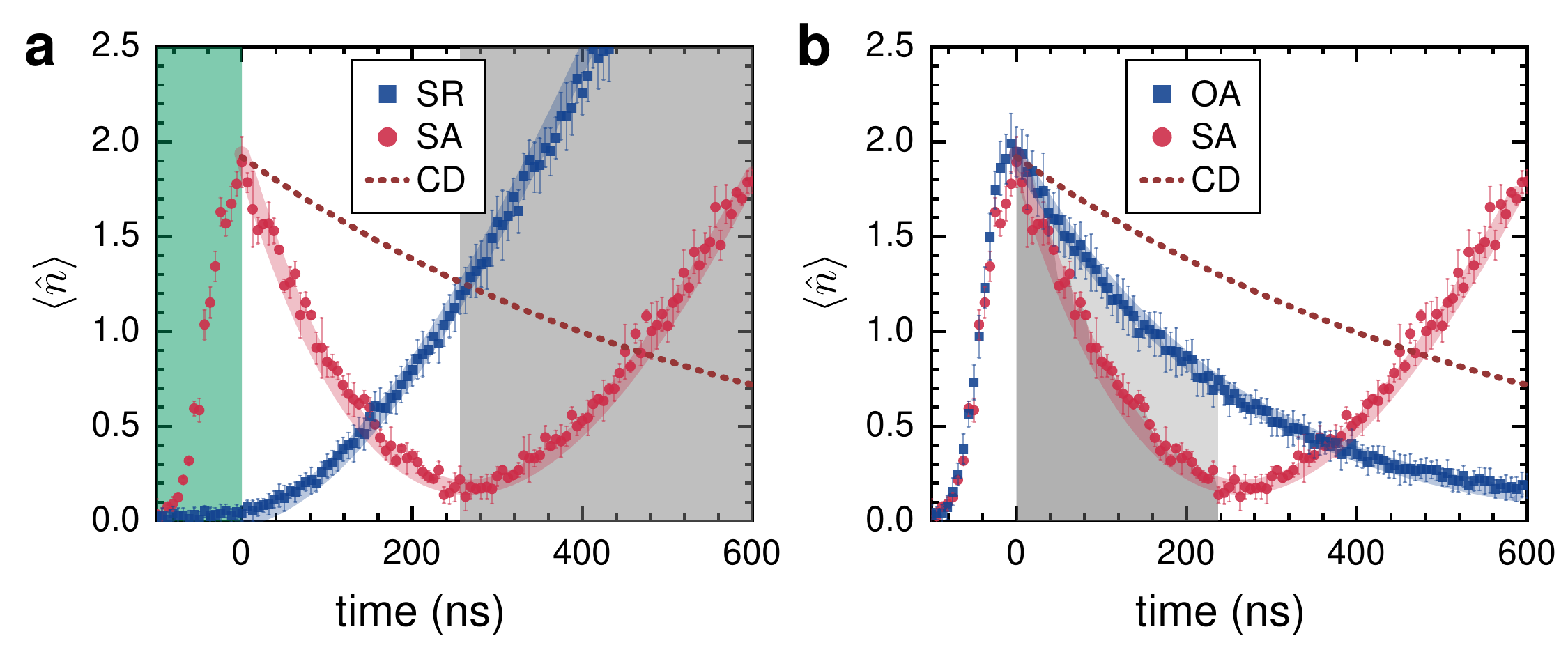}
	\caption{
		\textbf{Time-dependent intracavity photon numbers by superradiance, superabsorption and ordinary absorption. 
		}
		{\bf a},
		Starting from zero, the mean photon number grows quadratically in time in superradiance (blue dots). With an opposite-phase field (in green-shaded region) prepared in the cavity initially, the superradiant state exhibits superabsorption (SA, red dots in the unshaded area) instead. After fully absorbing the photons, the superradiant state starts to radiate (grey-shaded region).  Mild asymmetry in the red-dot curve is due to the cavity decay (CD).
		{\bf b}, The superabsorption (red dots) of the superradiant state gives much faster absorption than the ordinary absorption (blue dots, labeled as OA) of the ground state of atoms. The mean number of atoms in the cavity is the same for both cases ($N\approx6.8$). Thick-shaded blue and red curves in {\bf a} and {\bf b} represent the theoretically calculated mean photon numbers with no adjustable parameters. Error bars indicate the standard deviations from repeated measurements.	
%		\comment{[150words]}
		}
\label{fig:SR/SA/GND}
\end{figure}

In order to investigate the aforementioned time-reversal relation between the superradiance and the superabsorption, we observed both phenomena, radiation and absorption process of a superradiant state in a time-dependent manner(Fig.~\ref{fig:SR/SA/GND}{\bf a}). With the cavity initially empty, we let the superradiant-state atoms go through the cavity and emit photons into the cavity mode. During the photon buildup, 
the photon number increases quadratically in time (blue-dot curve in Fig.~\ref{fig:SR/SA/GND}{\bf a}) because the amplitude of the electric field generated by the constant macro dipole increases linearly in time in the slowly varying envelop approximation. Specifically, the mean number of photons is given by $\expval{\hat{n}(t)} \simeq |\rho_{\rm eg}Ng t |^2$ (see Supplementary Note 2 for derivation) with $\rho_{\rm eg}$ the off-diagonal element of the atomic density matrix. However, if we initially prepare the cavity with the coherent state $\ket{-\alpha}_{\rm f}$, the superradiant-state atoms absorb the photons in the cavity instead of emitting photons, undergoing a time-reversed process of superradiance (red-dot curve in Fig.~\ref{fig:SR/SA/GND}{\bf a}). After fully absorbing photons in the cavity, the superradiant state starts to radiate 
(see the grey-shaded region in Fig.~\ref{fig:SR/SA/GND}{\bf a} and Eq.~(\ref{eq5}) in Methods). The temporal profile of the radiation and that of the absorption are nearly symmetric with a mild imbalance due to the cavity decay. 
The ratio of the signals at equal time interval $t_0$ from the minimum point is approximately given by $e^{-2\gamma_c t_0}$ (see Supplementary Note 9).

For the enhanced absorption by the superradiant-state atoms to be called ``super'' absorption, the absorption rate of the superradiant state should be much faster than that of the ground state. In Fig.~\ref{fig:SR/SA/GND}{\bf b}, the superradiant state atoms clearly give faster absorption than the ground state atoms. Since a part of photons leaks out due to the cavity decay, for a fair comparison we define the absorption ratio $r_{\rm ab}$ as the number of {\it absorbed} photons for a given time interval $t_{\rm ab}$ to the total number of photons initially prepared in the cavity. We can calculate the number of cavity-decayed photons by numerically integrating the expression $2\gamma_{\rm c}\int_{0}^{t_{\rm ab}} \expval{\hat{n}(t)} dt$. Taking the numbers of cavity-decayed photons into account, with the integral corresponding to the grey shaded areas in Fig.~\ref{fig:SR/SA/GND}{\bf b}, we obtain the absorption ratios $r_{\rm SA}=0.85\pm0.07$ for the superabsorption (SA) and $r_{\rm OA}=0.37\pm0.07$ for the ordinary absorption (OA). Simple algebra (see Supplementary Note 3) shows we need 5.2 times more atoms for the ordinary absorption in order to achieve the same absorption ratio as the superabsorption in Fig.~\ref{fig:SR/SA/GND}{\bf b}.

\begin{figure}[t]
\includegraphics[width=0.5\textwidth]{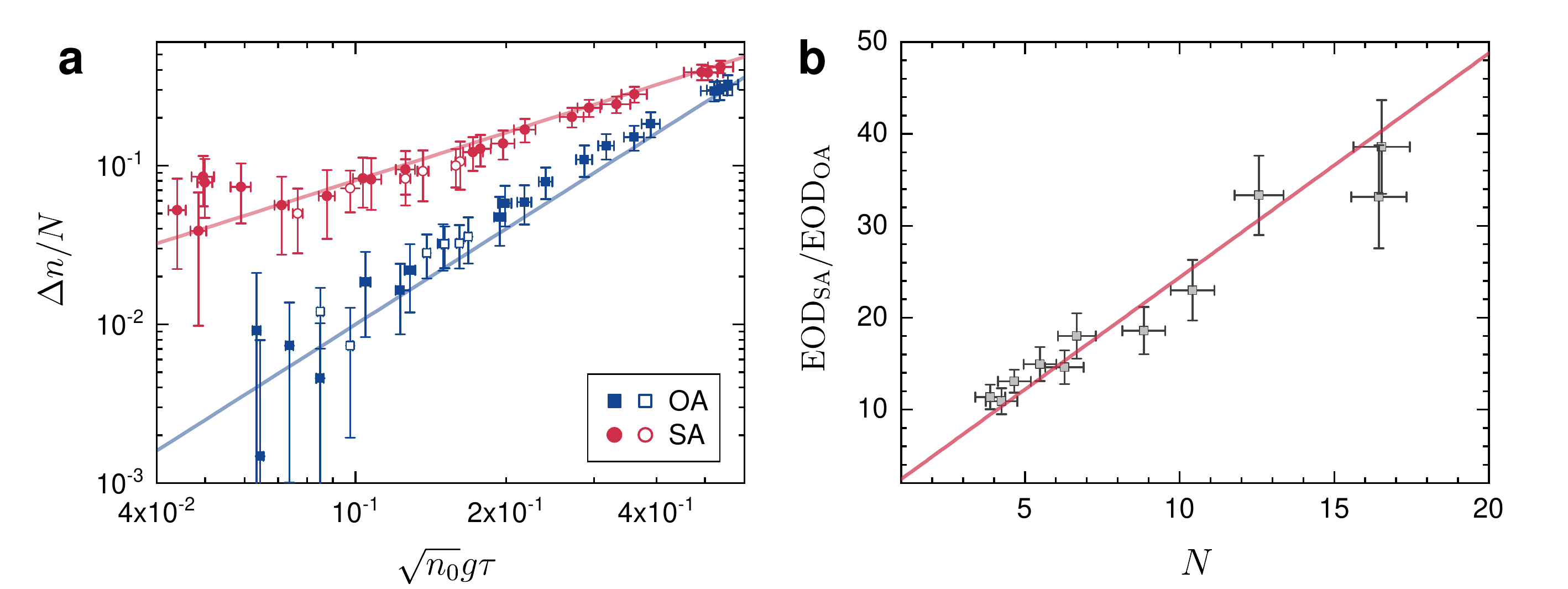}
	\caption{
	{\bf Comparison of superabsorption and ordinary absorption.} {\bf a}, The number of the absorbed photons per atom for a time interval $\Delta t=\tau$ in superabsorption (SA) as well as in ordinary absorption (OA).  Solid symbols represent experimental data with $g\tau$=0.16 and hollow ones correspond to the experiments with $g\tau$=0.10. Solid lines are theoretical expectations given by Eqs.\ (S20a) and (S20b) in Supplementary Note 5, respectively. {\bf b}, Ratio of the effective optical depths as a function of the mean number $N$ of atoms. Experimental data are denoted as squares and the solid line represents the approximation by Eq.\ (S19) in Supplementary Note 4. The approximation is valid for $N\gg 1$. The initial photon number was $n_0=1.0$. The effective optical depth for superabsorption is much larger than that of the ordinary absorption and the ratio scales with $N$. Error bars indicate standard deviations from repeated measurements. 
%	\comment{[140words]}
}
    \label{fig:absorption_enhancement}
\end{figure}

It can be shown that the superabsorption rate is always larger than the ordinary absorption rate. This inequality can be easily understood by considering $\Delta n/N$, the number of absorbed photons per atom, in both cases for a short time interval of $\Delta t$.
The quantity is approximately given by
$(\Delta n / N)_{\rm SA}=
\sqrt{n_0}g\Delta t$ and $(\Delta n / N)_{\rm OA}=n_0 g^2\tau\Delta t$ with $n_0$ the initial mean number of photons (see Supplementary Note 5 for derivations). Therefore, we have the ratio $\frac{(\Delta n / N)_{\rm OA}}{(\Delta n / N)_{\rm SA}}=\sqrt{n_0}g\tau=(\Delta n / N)_{\rm SA}|_{\Delta t=\tau}$. This ratio is always less than unity since at most one photon can be absorbed by a two-level atom without re-emitting photons. 
In the regime of usual absorption with a dephasing rate ($\sim 1/\tau$ in our case) much larger than the radiative decay rate ($\sim 2\gamma_{\rm a}$), we get $\sqrt{n_0}g\tau \ll 1$ (see Supplementary Note 5) and thus the superabsorption rate becomes much greater than the ordinary absorption rate. 
This consideration is supported by the data shown in Fig.~\ref{fig:absorption_enhancement}{\bf a}, where observed $(\Delta n / N)_{\rm SA}$ and $(\Delta n / N)_{\rm OA}$ are compared. The maximum number of absorbed photons per atom was as large as 0.4 for superabsorption, approaching the theoretical upper bound of 0.5 of the present scheme.
One can also notice that the superabsorption would excel for a small number of initial photons, implying usefulness in faint-light sensing\cite{Reiserer-science03}.

In the superabsorption, a coherent input field can be completely absorbed by the atoms in a finite time given by $t_0\simeq\frac{\sqrt{n_0}}{|\rho_{\rm eg}|N g}$ for a given initial mean photon number $n_0$. In contrast, such complete absorption cannot be achieved in a finite time in the ordinary absorption exhibiting an exponential decay of the input field. In the superabsorption data of Fig.~\ref{fig:SR/SA/GND}, a small number of photons($\expval{\hat{n}}\sim 0.2$) remained unabsorbed. These photons are in part incoherent photons generated by atomic spontaneous emission due to nonzero excited-state population $\rho_{\rm ee}$. Experimental imperfections such as incomplete phase matching between the input field and the atomic dipoles as well as the phase variation among the dipoles due to the finite nanohole size also contribute to the residual photons. Except for such experimental imperfections, the degree of absorption for superabsorption as well as ordinary absorption can be characterised by considering the effective optical depth (EOD)\cite{CEAS03}, an equivalent free-space optical depth, which in our case is given by ${\rm EOD}=\frac{r_{\rm ab}}{1-r_{\rm ab}}$ (see Supplementary Note 4). It equals the ratio of the number of absorbed photons to that of unabsorbed photons (including cavity-decayed photons). It can be shown analytically (see Supplementary Note 4) that the EOD for superabsorption is much larger than that of the ordinary absorption, by a factor proportional to the mean number of atoms. This feature is experimentally confirmed in Fig.~\ref{fig:absorption_enhancement}{\bf b}, where the EOD for superabsorption is measured to be at most 40 times larger than that of the ordinary absorption.

\begin{figure}
\includegraphics[width=0.475\textwidth]{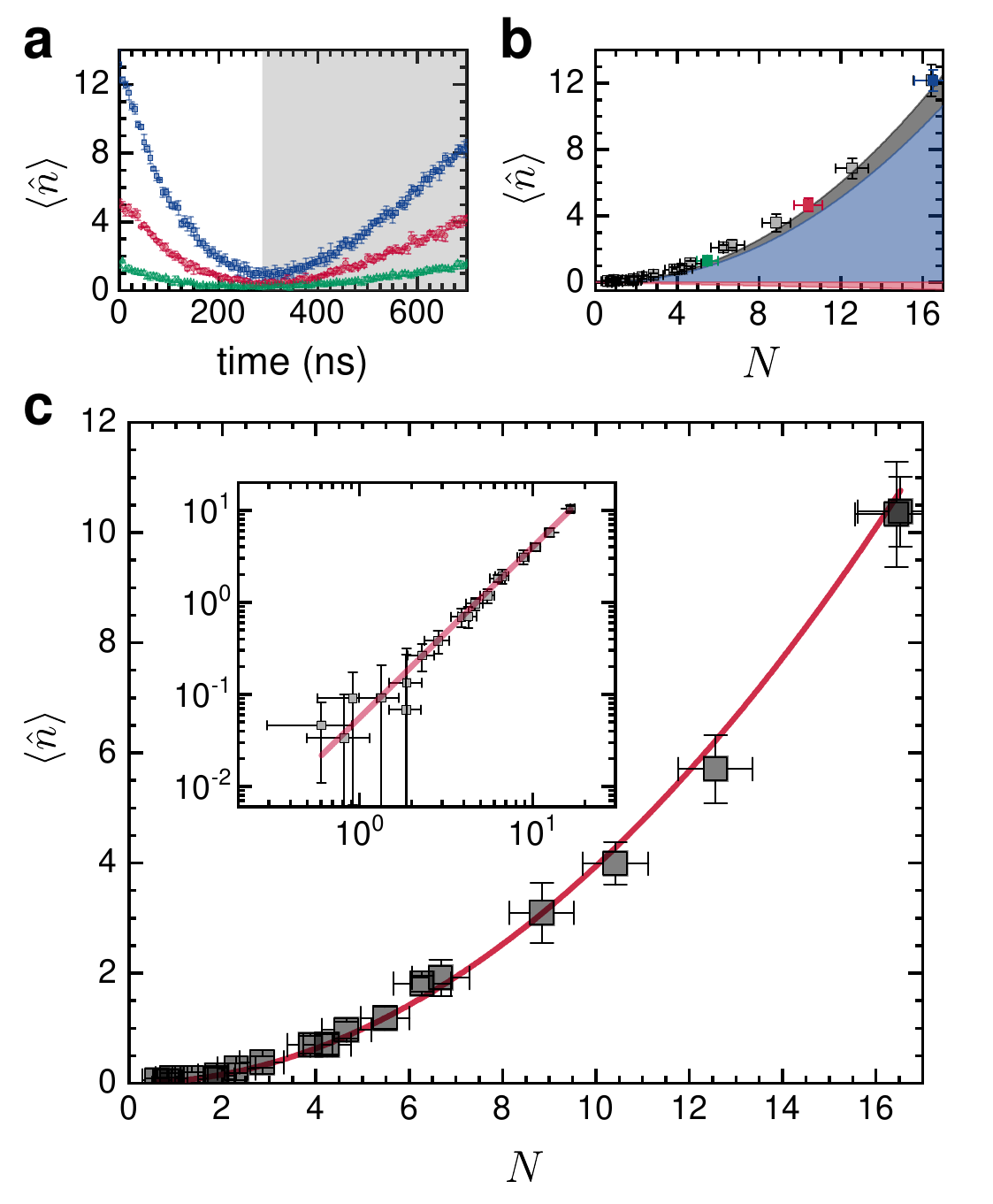}
	\caption{
		\textbf{{Quadratic}  dependence  of  superabsorption  on  atom  number. }
		{\bf a}, For a given number of atoms in the cavity, the maximum number of photons that can be completely absorbed is evaluated in a fixed time interval ${t_{0}}$. 
		{\bf b}, The reduced number of photons for the time interval from $t=0$ to ${t_{0}}$(=280ns) in {\bf a} 
		versus the mean number $N$ of atoms in the cavity. The upper grey-shaded area indicates the theoretically expected photon number reduction from the cavity decay while the lower red-shaded area indicates the increase by the spontaneous emission. The absorption by the correlated dipoles then corresponds to the sum of the blue- and the red-shaded areas.
		{\bf c}, The absorbed number of photons obtained in {\bf b} versus $N$. The $N$-dependence is well fit by $N^q$ (red curve) with $q=1.86\pm0.03$. The inset shows a linear fit in the log-log plot. Error bars indicate the standard deviations from repeated measurements. 
%		\comment{[148words]}
		}\label{fig:Nsq}
\end{figure}

One of the most distinctive features of superradiance is its nonlinear scaling of radiation intensity on the number of atoms, that is, the maximum number of emitted photons in a fixed time interval is proportional to $N^2$. Likewise, we can consider nonlinear scaling of the absorbed photons in superabsorption(Fig.~\ref{fig:Nsq}). We measured the maximum number of photons that can be completely absorbed in a fixed time interval as a function of the mean number of atoms(Fig.~\ref{fig:Nsq}{\bf a}). Note that in the actual experiment the photon number is reduced by both the cavity decay and the absorption by the correlated atoms but increased by the atomic spontaneous emission. Their individual contributions calculated from the theory are indicated in Fig.~\ref{fig:Nsq}{\bf b}. 
The number of purely absorbed photons by the correlated atoms are plotted in the log-log scale in Fig.~\ref{fig:Nsq}{\bf c}. The result can be well fit by the expression $\expval{\hat{n}}\propto N^q$ with $q=1.86\pm 0.03$. The small deviation from $N^{2}$ can be explained by the aforementioned experimental imperfections.

In summary, we have experimentally realised superabsorption by implementing time-reversed process of coherent superradiance with correlated atoms in a high-Q cavity. 
It is observed that the rate of superabsorption is much higher than that of ordinary ground-state absorption. The number of absorbed photons scales with the square of the number of atoms as a result of atomic cooperative interaction with the input field.
By noting that our scheme described by Eqs.~(\ref{eq:SR}) and (\ref{eq:SA}) does not mandate a cavity, we expect that superabsorption can also be possible by employing atoms in a superposition state interacting with a tightly focused laser field in free space. 
It might be possible to employ two-level systems with much faster transition rates along with frequency- and time-division multiplexing of continuous-wave input light to make our scheme more relevant to practical application.
Since the quantum state of absorbed photon is {\it coherently} mapped to the atomic internal states, our superabsorption can also be utilised in coherence-critical applications such as quantum memory\cite{QI_Quantum_memory_nat04} and light-matter quantum  interfaces\cite{QI_quantum_interface_rmp10}.\\

\noindent {\large \bf Methods}

%\comment{[Subsection on measurement procedure is moved to SM because Methods cannot have a figure in it. The remaining subsections are rearranged.]}

\noindent{\bf Superabsorption by reversing superradiance in time. }
The atom-field interaction in our system is described by the Tavis-Cummings Hamiltonian,
\begin{equation}\label{eq:TC_Hamiltonian}
	\hat{H}=\hbar g \sum_{i}^{N} \left( \hat{a}^\dagger \hat{\sigma_i} + \hat{a} \hat{\sigma_i} ^\dagger \right),
\end{equation}
where $g$ is the atom-cavity coupling constant, $\hat{a}(\hat{a}^\dagger)$ is the annihilation(creation) operator for the cavity field and $\hat{\sigma_i}(\hat{\sigma_i}^\dagger)$ is the lowering(raising) operator for the $i$th atom.
Superradiance process of a superradiant state $\ket{\Psi}_{\rm a}$ is then approximately described by the following time evolution operation.
\begin{equation}\label{eq:SR2}
	\hat{U}(t)\ket{\Psi}_{\rm a}\ket{0}_{\rm f} = \ket{\Psi'}_{\rm a}\ket{\alpha}_{\rm f},
\end{equation}
where $\hat{U}(t)\equiv e^{-i\hat{H}t/\hbar}$, $\ket{0}_{\rm f}$ denotes a photonic vacuum state, $\ket{\alpha}_{\rm f}$ denotes a photonic coherent state with an amplitude $\alpha$, and $\ket{\Psi'}_{\rm a}$ denotes the resulting atomic state by the time evolution. We introduce a field-phase flip operator $\hat{R}_{\pi}=e^{-i\pi \hat{a}^\dagger\hat{a}}$, corresponding to $\pi$-rotation in the field phase space\cite{flip-operator}. The operator $\hat{R}_{\pi}$ only {acts} on the photonic state and satisfies the property, $\hat{R}_{\pi}\hat{a}\hat{R}_{\pi}^{\dagger} = -\hat{a}$ and similarly $\hat{R}_{\pi}\hat{a}^\dagger\hat{R}_{\pi}^{\dagger}= -\hat{a}^\dagger$, and thus $ \hat{R}_{\pi}^\dagger\hat{U}(t) \hat{R}_{\pi}=\hat{U}(-t)$.

The result from simply applying $\hat{U}(-t)$ on both side of Eq.~(\ref{eq:SR2}) is the time-reversed process of superradiance, $\hat{U}(-t) \ket{\Psi'}_{\rm a}\ket{\alpha}_{\rm f}=\ket{\Psi}_{\rm a}\ket{0}_{\rm f}$, which is superabsorption. Since direct time-reversing is not experimentally possible, we utilise a phase-flip operation. Let us consider
\begin{eqnarray} \label{eq:SR_Time-reversed_detail}
	&\hat{U}(t')&\ket{\Psi'}_{\rm a}\ket{-\alpha}_{\rm f}=\hat{U}(t')\hat{R}_{\pi}^\dagger\hat{R}_{\pi}\ket{\Psi'}_{\rm a}\ket{-\alpha}_{\rm f}\nonumber\\
	&=&\hat{U}(t')\hat{R}_{\pi}^\dagger\ket{\Psi'}_{\rm a}\ket{\alpha}_{\rm f}=\hat{R}_{\pi}^\dagger\hat{U}(-t')\ket{\Psi'}_{\rm a}\ket{\alpha}_{\rm f}\nonumber\\
	&=&\hat{R}_{\pi}^\dagger\hat{U}(-t'+t)\hat{U}(-t)\ket{\Psi'}_{\rm a}\ket{\alpha}_{\rm f}\nonumber\\
	&=&\hat{R}_{\pi}^\dagger\hat{U}(-t'+t)\ket{\Psi}_{\rm a}\ket{0}_{\rm f}=\hat{U}(t'-t)\hat{R}_{\pi}^\dagger\ket{\Psi}_{\rm a}\ket{0}_{\rm f}\nonumber\\
	&=&\hat{U}(t'-t)\ket{\Psi}_{\rm a}\ket{0}_{\rm f}\;.
\label{eq5}
\end{eqnarray}
If $t'=t$, we completely reverse the superradiance to recover the initial vacuum state. This is superabsorption. If $t'<t$, the reversal is not complete. If $t'>t$, we then get a normal superradiance process progressing for a time period of $t'-t$. Therefore, we can realise time-reversal process of superradiance by preparing the field state as $\ket{-\alpha}_{\rm f}$ and by letting the system evolve in an appropriate time.\\

\noindent{\bf Nanohole-array aperture. }
The nanohole-array aperture was fabricated by using the focused ion beam technique on a 10nm-thick silicon nitride membrane mounted on a 200$\mu$m-thick silicon frame. The individual hole size measured with a scanning-electron microscope is $0.35\lambda\times0.24\lambda$ ($\lambda$=791nm). The horizontal spread of the atomic distribution by the finite hole size has an effect of averaging the atom-cavity coupling. The resulting variation in $g$ is $\Delta g/g=0.20$. The vertical spread of the atomic distribution decreases atomic phase purity due to the position dependence of the pump laser phase. The height of the hole was thus intentionally made smaller than the width in order to minimise atomic phase variation while maintaining an enough atomic flux. The aperture is mounted on a 6-axis stage with an additional piezoelectric transducer(PZT) for precise alignment along the cavity axis. The aperture is coarsely placed at the center of the waist of the cavity mode (vertically and horizontally) by the 6-axis stage and then a feedback loop places the center of each hole at cavity anti-nodes by using the PZT. The average atom-cavity coupling is decreased additionally because the vertical size (40$\mu$m) of the nanohole-array aperture is comparable to the cavity mode waist(43$\mu$m), resulting in $\Delta g/g=0.06$.\\

\noindent{\bf Atom Number Measurement.}
To measure the mean number of atoms in the cavity, $^1{\rm S}_0 \leftrightarrow ^1{\rm P}_1$ transition is used. The wavelength of the transition is 553nm with a radiative decay rate of 18.9MHz. While illuminating the atoms in the cavity with a resonant probe laser propagating along the cavity mode, the fluorescence photons were collected from the side.
The mean number of atoms in the cavity $N$ and the mean number of photons in the cavity $\expval{\hat{n}}$ are simultaneously calibrated by measuring both quantities in the nonlinear $\expval{\hat{n}}$-vs.-$N$ regime of the conventional microlaser lasing and by comparing the data with the well-known master equation theory\cite{ML_SR}. The number of atoms is controlled by a mechanical `blocker', partially blocking the atomic beam flux well in front of the nanohole array aperture as well as the cavity. Although the blocker is supposed to change the mean atom number with a step size smaller than 0.1, the precision of atomic fluorescence measurement is not high enough to resolve such stepwise changes.\\

\newpage
%\hbox{}

%\hbox{}
\noindent\textbf{Data availability}

\noindent
%Source data for Figs.~1-4 and Fig.~S1 are provided. The data that support the findings of this study are available from the corresponding author upon reasonable request.\\
The data that support the plots within this paper and other findings of this study are available from the corresponding author upon reasonable request.\\

\noindent\textbf{Code availability}

\noindent
All data are obtained from the experiments or from analytic formulae discussed in the manuscript. No special computer codes are used to generate the results reported in this paper.\\

%\comment{[Reference for flipping operator is added as [30]]}

%\bibliographystyle{naturemag}
%\bibliography{SA_ref-latest}

\vspace{0.15in}

\noindent\textbf{Acknowledgements}

\noindent
This work was supported by Samsung Science and Technology Foundation (Project No. SSTF-BA1502-05), the Korea Research Foundation (Grant No.~2020R1A2C3009299) and the Ministry of Science and ICT of Korea under ITRC program (Grand No.~IITP-2019-2018-0-01402).\\

\noindent\textbf{Author Contributions}

\noindent
D.Y. and K.A. conceived the experiment. D.Y. performed experiments with help
from S.O., J.H. and G.S.. D.Y. analysed the data and carried out theoretical investigations with help from Ju.K.. K.A. supervised overall experimental and theoretical works. D.Y. and K.A. wrote the manuscript. All authors participated in discussions.\\

\noindent\textbf{Competing Interests}

\noindent
The authors declare no competing financial interests.\\ 

\noindent\textbf{Author Information}

\noindent
Correspondence and requests for materials should be addressed to K.A. (kwan@phya.snu.ac.kr).

\end{document}